\documentclass[lettersize,journal]{IEEEtran}

\usepackage[noadjust]{cite}
\usepackage[english]{babel}
\usepackage{blindtext}
\usepackage{booktabs} 
\usepackage{epsfig,endnotes, enumerate}
\usepackage{subfigure}
\usepackage{graphicx}
\usepackage{url}
\usepackage{color}
\usepackage{balance}
\usepackage{amssymb,amsmath, listings, verbatim, float}
\usepackage[linesnumbered, ruled, vlined]{algorithm2e}
\usepackage{mathtools, comment}
\usepackage{multirow}
\usepackage{etoolbox}
\usepackage{enumitem}

\definecolor{LightCyan}{rgb}{0.0,1,1}
\definecolor{LightGray}{rgb}{0.8,0.8,0.8}

\definecolor{blue}{rgb}{0.0,0.0,1}
\newcommand{\revise}{\textcolor{black}}

\definecolor{red}{rgb}{1,0.0,0.0}

\definecolor{gray}{rgb}{0.3,0.3,0.3}
\newcommand{\code}[1]{\textcolor{gray}{\texttt{#1}}}

\begin{document}

\title{Enabling Scalability in Asynchronous and Bidirectional Communication in LPWAN}


\author{Mahbubur Rahman\\
Computer Science, Graduate Center and Queens College\\
City University of New York
\thanks{Mahbubur Rahman is with the City University of New York--Graduate Center and Queens College. This manuscript is an extended version our conference paper published in IEEE ICESS '22~\cite{rahman2022enabling}.}}


\maketitle

\begin{abstract}
\revise{Low-power wide-area networks (LPWANs) have become ubiquitous in the Internet of Things (IoT) applications due to their ability to connect sensors over large geographic areas in a single hop. It is, however, very challenging to achieve massive scalability in LPWANs, where numerous sensors can transmit data efficiently and with low latency, which emerging IoT and CPS (cyber-physical systems) applications may require. In this paper, we address the above challenges by significantly advancing an LPWAN technology called SNOW (sensor network over white spaces). SNOW exploits distributed orthogonal frequency division multiplexing (D-OFDM) subcarriers to enable parallel reception of data to a base station (BS) from multiple asynchronous sensors, each using a different subcarrier. 
In this paper, we achieve massive scalability in SNOW by enabling the BS to decode concurrent data from numerous asynchronous sensors on the same subcarrier while parallelly decoding from other subcarriers as well. Additionally, we enable numerous asynchronous sensors to receive distinct data from the BS on the same subcarrier while other sensors also receive data parallelly on other subcarriers. To do this, we develop a set of Gold code-based pseudorandom noise (PN) sequences that are mutually non-interfering within and across the subcarriers. Each sensor uses its PN sequence from the set for encoding or decoding data on its subcarriers, enabling massive concurrency. Our evaluation results demonstrate that we can achieve approximately 9x more scalability in SNOW while being timely in data collection at the BS and energy efficient at the sensors. This may enable emerging IoT and CPS applications requiring tens of thousands of sensors with longer battery life and making data-driven, time-sensitive decisions.}

\end{abstract}


\section{Introduction}\label{sec:intro}
\begin{figure}[!htpb]
\centering \vspace{0.04in}
\includegraphics[width=0.48\textwidth]{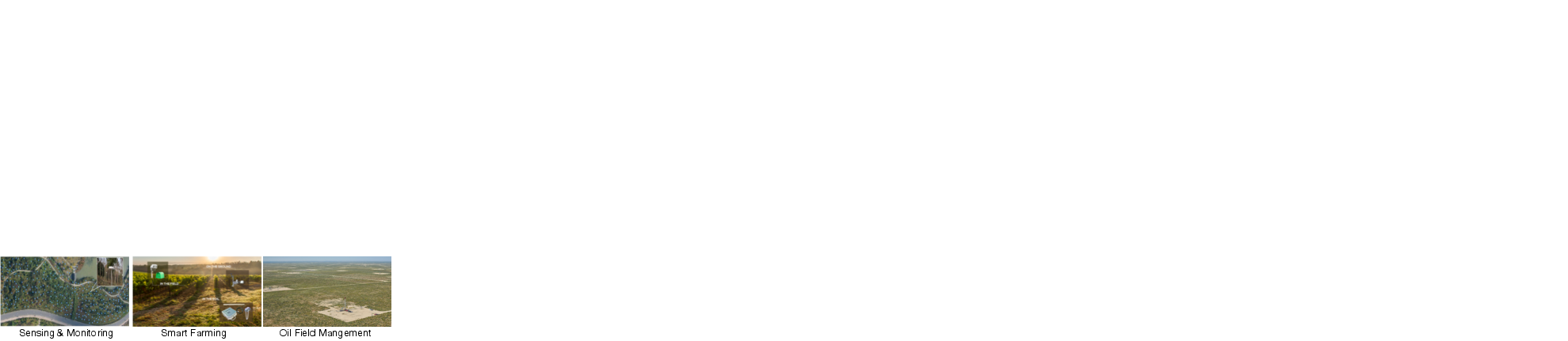}
\caption{A few examples of emerging IoT and CPS applications, demonstrating their extensive geocoverage and significant scalability requirements.}\vspace{-0.1in}
\label{fig:apps}\vspace{-0.1in}
\end{figure}

The number of Internet of Things (IoT) connections is expected to reach 40 billions by the year 2030, with an industry value of over a trillion dollars. The emerging IoT and CPS (cyber-physical systems) applications, including sensing and monitoring, smart farming, and oil field management aim to utilize IoT devices for enhancing sustainability, quality of life, health, safety, and economic prosperity of communities in both urban and rural areas. IoT devices (i.e., sensors or simply nodes) are usually battery-powered, scattered in large numbers (e.g., tens of thousands) over vast and various distances (e.g., tens of kilometers) for the above use cases (see Figure~\ref{fig:apps} as a reference). It thus becomes extremely challenging to connect and coordinate these sensors for periodic or sporadic data collection and make time-critical, data-driven decisions.

To enable wide-area IoT and CPS applications, existing wireless sensor network (WSN) technologies, including Zigbee and WirelessHART form multi-hop mesh networks, complicating the protocol design and network deployment resulting in scalability issues in applications, high energy consumption at the nodes, and high latency in data collection at the BSs~\cite{survey_wsn}. Due to their underlying design and operational limitations, existing low-power wide-area network (LPWAN) technologies, including LoRa, SigFox, NB-IoT, and 5G also suffer from scalability issues, high energy consumption, and high latency in sensor data collection, especially in infrastructure-limited rural areas~\cite{survey_lpwan}. For example, the leading LPWAN technology, LoRa, supports approximately 120 nodes per 3.8 hectors until its performance drops sharply~\cite{bor2016lora}, which may not be sufficient to meet the scalability and sustainability requirements of the emerging IoT and CPS applications~\cite{survey_lpwan, rahman2023boosting}.

To this extent, we focus on enabling massive scalability in an LPWAN technology called SNOW (sensor network over white spaces)~\cite{saifullah2016snow, saifullah2017enabling, saifullah2018low, rahman2019implementation}. The current SNOW design exploits the {\em TV white spaces} -- allocated but locally unused TV channels that can be used by unlicensed devices~\cite{fcc_second_order}) -- to connect sensors to a BS. SNOW has a D-OFDM (distributed orthogonal frequency-division multiplexing) based physical (PHY) layer that allows different {\em asynchronous} sensors (e.g., need no coordination needed between sensors) to transmit data {\em concurrently} to a BS in uplink communications using different D-OFDM subcarriers or subchannels~\cite{saifullah2016snow}. D-OFDM also allows the SNOW BS to transmit distinct data to different sensors in downlink communications both asynchronously and concurrently using different subcarriers~\cite{saifullah2017enabling}. 
\revise{When the number of subcarriers is fewer than the number of nodes, SNOW allocates the same subcarrier to multiple nodes for both uplink and downlink communications. In such a scenario, the sensors operating in the same subcarrier employ carrier sensing in the uplink communications, which results in higher energy consumption at the sensors and increased latency in convergecast at the BS, particularly exacerbated by the hidden terminals in the network. In downlink communications, the nodes operating in the same subcarrier receive data in a round-robin fashion set by the BS during the subcarrier assignment phase through a unique (set of) join-subcarrier.}

\revise{In this paper, we enable massive scalability, higher energy efficiency, and decreased latency in both uplink and downlink communications in SNOW as follows. 
\\
\indent {\bf (1)} We enable numerous {\em asynchronous} sensors to {\em concurrently} transmit data using the same subcarrier to the BS, while, in parallel, the other sensors may also transmit using the rest of the subcarriers in a similar fashion.
\\
\indent {\bf (2)} We enable the BS to transmits distinct data to numerous {\em asynchronous} sensors listening {\em concurrently} to the same subcarrier, while, in parallel, the other sensors may also listen to and receive in a similar fashion on the other subcarriers.}

\noindent \revise{Enabling such massive concurrency in SNOW uplink and downlink communications is, however, a very challenging task, particularly for the following reasons.
{\bf First}, concurrent transmissions from two or more sensors on the same subcarrier result in a typical {\em collision} scenario, which makes it impossible for the BS to decode any of the transmissions. This also results in lost packets in the network, wasted energy consumption at the sensors, and increased latency in convergecast. 
{\bf Second}, parallel transmissions from different sensors on the neighboring (i.e., adjacent) subcarriers break the {\em orthogonality} of the D-OFDM architecture, which also makes it impossible for the BS to decode any of the transmissions, resulting in similar consequences to those in the first case.
In a nutshell, the above two challenges introduce severe {\em inter-symbol} interference between the signals transmitted from the sensors on the same subcarrier and {\em inter-subcarrier} interference between the signals transmitted from sensors on the neighboring subcarriers.
In the case of downlink communications from the BS to the sensors, these challenges also plague the SNOW D-OFDM architecture, which results in reduced performance.}

To this extent, we address the above challenges and make the following key contributions.
\begin{itemize}[topsep=0pt]


  \item \revise{We develop a set of decentralized {\em pseudorandom noise} (PN) sequences (a.k.a. {\em pseudorandom spreading sequence}) based on Gold code~\cite{gold1967optimal}. These PN sequences have very good {\em cross-correlation} properties, e.g., the correlation value between any pair of PN sequences in the set is minimal or close to zero, making them orthogonal to each other on and across the D-OFDM subcarriers.}

  \item \revise{We enable concurrent transmissions on the same D-OFDM subcarrier from (to) asynchronous sensors to (from) the BS by assigning each sensor a different sequence from the same set of PN sequences, which mitigates the inter-symbol interference within that subcarrier. To minimize the inter-subcarrier interference, we assign the sensors operating on the neighboring subcarriers (on both sides) distinct PN sequences from another set of PN sequences generated using different {\em seeds} while maintaining the required cross-correlation properties with the earlier set of PN sequences.}

  \item We enable a higher bitrate than the per-sensor bitrate requirement of the IEEE 802.15.4 standards'~\cite{ieee802154} direct-sequence spread spectrum (DSSS) that spreads a group of 4 bits to 32 chirps, considering a typical sensor data size of 28 bytes in practical deployments (e.g., for those using TinyOS~\cite{tinyos}). Our design may thus inspire enhanced scalability in the WSN standards as well.

  \item \revise{Additionally, we provide a numerical scalability analysis of our design and compare with it LoRa (the industry-leading LPWAN technology) and Sigfox. Our analysis shows that our design may provide {\em significantly} higher scalability in emerging IoT and CPS applications, which may inspire the IoT industry to focus on SNOW as well.}

  \item Finally, we develop a SNOW simulation platform using Python's NumPy library and make it open-source~\cite{snowsimulation}. In simulation, we implement the SNOW PHY layer, including our innovations, and perform a large-scale evaluation. Our evaluation results show that our design may provide approximately 9x improvements in scalability compared to the existing SNOW design, resulting in better energy efficiency in the sensors and reduced latency in data collection at a BS in convergecast scenarios.
\end{itemize}

\revise{In rest of the paper, Section~\ref{sec:related} presents the related work. Section~\ref{sec:overview} briefly overviews the existing SNOW architecture and presents our system model. Section~\ref{sec:technical} details our PN sequences generation techniques for spreading and despreading data and analyses on achievable bitrate and scalability. Section~\ref{sec:eval} provides the implementation details and evaluation results. Finally, Section~\ref{sec:conclusion} concludes our paper.}

\section{Related Work}\label{sec:related}
In this section, we provide a comparative study between SNOW and the other contemporary wireless technologies.

\subsection{WSN Technologies} 
The emerging wide-area IoT and CPS applications need to connect and coordinate hundreds to thousands of sensors over distances of tens of kilometers. The existing WSN technologies operating in the 2.4 GHz spectrum (e.g., IEEE 802.15.4, IEEE 802.11, and BLE) may facilitate such connections by forming multi-hop mesh networks due to their short communication range~\cite{survey_wsn, kim2007health}.
This, however, will complicate the protocol design, resulting in reduced scalability, high energy consumption at the sensors, high latency in data aggregation, and high cost in real-world deployments~\cite{rahman2021lpwan, saifullah2018low, modekurthy2023towards}. In this paper, we develop protocols for enhanced scalability in LPWANs that have the potential to connect numerous sensors to a BS by forming a single-hop over several kilometers.

\subsection{LoRa and Sigfox}\label{sec:related_lora}
Sigfox and LoRa are the two dominating LPWAN technologies 
operating in the {\em unlicensed ISM band}~\cite{survey_lpwan}.
Their devices adopt a 1\% or 0.1\% duty cycle requirement, making them less suitable for IoT or CPS applications with thousands of sensors or with real-time requirements~\cite{survey_tvws, bor2016lora, voigt2017mitigating, marcelis2020dare, liando2019known, fahmida2020long, rahman2023boosting, fahmida2023handling}.
Sigfox supports a datarate of 10 to 1,000 bps, and a device can send at most 140 12-byte messages (each takes 3 seconds) per day. LoRa employs different channel bandwidths (BWs) between 125 and 500 kHz, spreading factors (SFs) between 7 and 12, and coding rates between $\frac{4}{5}$ and $\frac{4}{8}$ to achieve scalability and different datarates. Using 125 kHz BW and SF of 10, a 12-byte payload in LoRa has an air time of 411.6 ms and bitrate of 980 bps. The higher the SF, the lower the bitrate in LoRa. This problem is exacerbated since large SFs are used more often~\cite{adelantado2017understanding}. Sigfox and LoRa may not be suitable for the emerging IoT and CPS applications requiring massive scale, high data rate, and ultra-low latency~\cite{survey_lpwan, survey_tvws}. Conversely, SNOW has the potential to achieve the above in the TV white spaces~\cite{rahman2019implementation}, and hence, it is our focus in this paper.




\subsection{SNOW vs. Other LPWANs} 
A number of LPWAN technologies, including NB-IoT~\cite{chen2017narrowband} and 5G~\cite{akpakwu2017survey} have targeted the cellular infrastructure and band. 
The 5G standard is currently under development. 
The NB-IoT specification froze at Release 13 of the 3GPP specification.
Operating in the licensed band is costly due to high service fees and infrastructure and may not be available in the infrastructure-limited rural areas~\cite{survey_lpwan, vasisht2017farmbeats, chakraborty2022whisper}. These technologies also require the sensors to frequently synchronize, which is much energy-consuming. It thus is impractical to ensure sustainability over an extended period, uninterrupted operation, and longevity of the emerging IoT and CPS applications. Many other technologies have been developed that operate in the licensed (e.g., LTE Cat M1 and EC-GSM-IoT) or unlicensed (e.g., INGENU, IQRF, Telensa, DASH7, Weightless-N/P, IEEE 802.11ah, IEEE 802.15.4k/g) bands~\cite{survey_tvws, survey_lpwan, kouvelas2020p, hsieh2018experimental, saxena2016achievable, ikpehai2018low} and severely interfere each other (as applicable).
To avoid the high cost of the licensed band and the crowd of the ISM band, SNOW has been developed~\cite{modekurthy2024extending, rahman2022enabling, rahman2021lpwan, rahman2020integrating, rahman2018integrating, rahman2019implementation, rahman2018demo, ismail2018demo, rahman2020low, modekurthy2021low, saifullah2018low, saifullah2017enabling, saifullah2016snow}.  White spaces are widely available in both urban and rural areas, are less crowded, and offer a wider spectrum compared to other available frequencies for LPWANs~\cite{bahl2009white, survey_lpwan, survey_tvws, saifullah2017enabling}. SNOW thus has huge potential, and we propose to significantly advance its PHY layer.
\section{Background and System Model}\label{sec:overview}
In this section, we briefly overview the SNOW technology and present our system model and assumptions.
\subsection{Overview of SNOW}\label{sec:snowdesign}
In the following, we provide a concise description of the SNOW architecture, physical layer, and MAC layer.
\begin{figure}[!htbp]
    \centering 
    \includegraphics[width=0.4\textwidth]{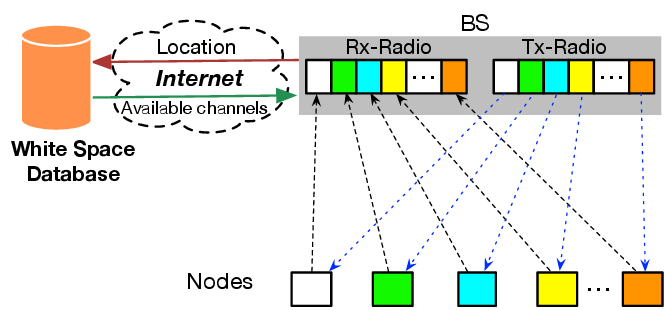} 
    \caption{SNOW Dual-radio BS and subcarriers.}
    \label{fig:dualradio}
\end{figure}
\subsubsection {Network Architecture}
SNOW is a an LPWAN technology operating in the TV white spaces. It supports asynchronous, reliable, bi-directional, and concurrent communication between a BS and numerous nodes. Due to its long-range, SNOW forms a star topology allowing the BS and the nodes to communicate directly, as shown in Figure~~\ref{fig:dualradio}. The BS is powerful, Internet-connected, and line-powered while the nodes are power-constrained and do not access the Internet. To determine white space availability in a region, the BS queries a cloud-hosted geolocation database. A node depends on the BS to learn its white space availability. In SNOW, all the complexities are offloaded to the BS to make the node design simple. Each node is equipped with a single half-duplex radio.

\subsubsection {Physical Layer}\label{sec:snowphy}
To support simultaneous uplink and downlink communications, the BS uses a dual-radio architecture for reception (Rx) and transmission (Tx), as shown in Figure~\ref{fig:dualradio}.
The SNOW PHY layer uses a distributed implementation of OFDM called {\em D-OFDM}. D-OFDM enables the BS to receive concurrent transmissions from {\em asynchronous} nodes using a single-antenna radio (Rx-radio). Also, using a single-antenna radio (Tx-Radio), the BS can transmit different data to different nodes concurrently. 
The BS operates on a wideband channel split into overlapping (50\%) orthogonal narrowband subchannels called subcarriers. Each node is assigned a subcarrier. 
For encoding and decoding on each subcarrier, the BS runs inverse fast Fourier transform (IFFT) and global fast Fourier transform (G-FFT) over the entire wideband channel, respectively. SNOW supports ASK (amplitude-shift-keying) and BPSK (binary phase-shift-keying) modulation techniques.


\subsubsection {Medium Access Control Layer}
When the number of nodes is no greater than the number of subcarriers, each node is assigned a unique subcarrier. Otherwise, a subcarrier is shared and the corresponding nodes use a lightweight CSMA/CA (carrier sense multiple access with collision avoidance)-based MAC (medium access control) protocol to uplink communication. The nodes can autonomously transmit, remain in receive mode, or sleep. When a node has data to send, it wakes up by turning its radio on. Then it performs a random back-off in a fixed initial back-off window. When the back-off timer expires, it runs CCA (clear channel assessment). If the subcarrier is clear, it transmits the data. If the subcarrier is occupied, then the node makes a random back-off in a fixed congestion back-off window. After this back-off expires, if the subcarrier is clean the node transmits immediately. This process is repeated until it makes the transmission and gets an acknowledgment (ACK). In downlink communication, the SNOW BS creates a round-robin schedule for the nodes operating on the same subcarrier to receive unique commands.

\subsection{Our System Model, Scope, and Assumptions}\label{sec:model1}
In the following, we describe our system model, scope of innovations in SNOW, and a few assumptions for this paper.
\begin{figure}[!htbp]
    \centering
    \includegraphics[width=0.32\textwidth]{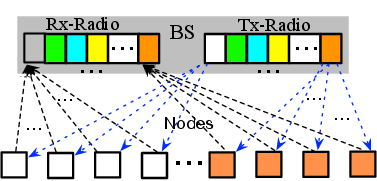} 
    \caption{Proposed uplink and downlink concurrency in SNOW PHY layer.}
    \label{fig:model}
\end{figure}

Currently, the SNOW BS can concurrently receive from or transmit to distinct sensors using only distinct subcarriers at any given instance (even with its MAC protocol), which limits the scalability in both uplink and downlink communications compared to its great potential. \revise{Additionally, its CSMA-based MAC protocol fails to account for hidden terminals. This may be caused by sensor-node mobility in the network or the BS assigning hidden sensors the same subcarrier involuntarily due to having a lesser number of subcarriers in its geolocation, further degrading the network scalability, energy consumption at the sensors, and latency in convergecast.}
\revise{In our design, we propose to enable concurrency in asynchronous sensors within and across the subcarriers at any given instance in both uplink and downlink communications, as depicted in Figure~\ref{fig:model}. Specifically, we enable the BS to decode data within a subcarrier from multiple sensors that do not coordinate themselves in time or frequency, while also decoding data from numerous other sensors in other subcarriers in parallel in a similar fashion.}

The proposed concurrency in our design increases the scalability of SNOW in both uplink and downlink by a factor of $\sum_{i=0}^{r-1}s_i$, where $s_i$ is the number of sensors using subcarrier $i$ concurrently and $r$ is the number of total subcarriers available.

In this paper, we limit our work to advancing the concurrency in uplink and downlink communications through the SNOW PHY layer only and leave the room for a new MAC protocol (needed when the number of nodes assigned to a subcarrier exceeds a subcarrier's concurrency capacity) as the future work. We thus solely focus on developing a set of PN sequences that preserves the D-OFDM feature of the SNOW PHY layer such that the inter-symbol interference and inter-subcarrier interference are minimal and the BS can decode data. Overall, the BS generates the set of PN sequences (and creates two instances of it) and assign each sensor a sequence when the sensor joins the network and assigned a subcarrier. No two sensors that are sharing a subcarrier or on neighboring subcarriers (on both sides) get the same PN sequence.
We also adopt many design parameters of the current SNOW architecture, including subcarrier overlapping (50\%), bandwidth (200--400 kHz), and subcarrier data modulation (e.g., ASK). \revise{To this extent, we refer to our design improvements as {\em mSNOW} and the current design as simply SNOW to avoid confusion. In the following, we now detail the design of mSNOW.}



\section{mSNOW: Enabling Massive SNOW PHY Layer Concurrency}\label{sec:technical}
\revise{In this section, we first detail our techniques for generating the set of PN sequences (i.e., pseudorandom spreading sequences). We then discuss our proposed encoding (i.e., spreading) of the transmitted signals and decoding (i.e., despreading) of the received signals in mSNOW for both uplink and downlink communications. Additionally, we discuss the achievable datarates and scalability by our proposed PHY layer innovations in mSNOW.}

\subsection{Spreading Sequence Preliminaries}
Recall that we enable concurrency within and across the D-OFDM subcarriers for any given instance. For this, we develop a set of PN sequences or waveforms, which allows numerous sensors to share a band of frequencies (i.e., subcarriers) with as little mutual interference as possible when each sensor is assigned a distinct sequence or code. Ideally, a received signal which has been spread using a different code will cause minimal interference in the aggregated signal over the entire bandwidth. The amount of interference from a sensor employing a distinct code (from a set) is related to the {\em cross-correlation} and power levels of all the codes in the set~\cite{kim1988system}. 
Unfortunately, such an ideal set would contain sequences of equally likely infinite random binary digits, requiring infinite storage in both the transmitter and receiver, and thus making impractical for the resource-constrained sensors.

The above limitations inspire the need for a set of {\em periodic} PN sequences (also used in Gold code~\cite{gold1967optimal}) that can be generated using a simplified circuit consisting of two {\em linear feedback shift registers} (LFSRs) and a few {\em XOR} (exclusive OR) gates (one for XORing two LFSRs and one for each tap in the LFSRs), which is practical for the sensors. The number of taps in each LFSR is determined by its unique polynomial equation~\cite{klein2013linear} and our achievable bitrate under minimum interference (explained in Section~\ref{sec:ourbitrate}). An LFSR generates {\em maximal-length sequences} (m-sequences) that are the pseudorandom binary sequences of the maximum period (e.g., until it repeats). An XOR gate is used to mix two different m-sequences (of the same length) from two different LFSRs to generate a PN sequence in our design.
In an LFSR, a bit is generated by a linear combination of the previous $n$ bits, for a suitable choice of $n$. In a nutshell, a window of $n$ bits (i.g., initial seed) is slide right (by one position) $2^n - 1$ times to cover $2^n - 1$ $n$-bit strings, generating $2^n - 1$ distinct m-sequences, each with a length of $2^n - 1$. We avoid $2^n$ slides since this starts repeating the sequences and \revise{may cause inter-symbol interference within a subcarrier and inter-subcarrier interference in neighboring subcarriers when the actual PN sequence is generated and used by the corresponding sensors.} In the following, we detail the m-sequences and our intended set of PN sequences generation techniques.

\subsection{m-Sequences Generation}
Each LFSR generates a maximum of $2^n - 1$ m-sequences, each of $2^n - 1$ bits, where $n$ is the number of bits in the initial seed~\cite{golomb1982shift}. Specifically, each LFSR register shifts all the bits to the right at each clock cycle (say, $c$), generating the $i$-th sequence (say, $a_i$), which may be represented using the following recursive equation~\cite{dinan1998spreading}:
$$a_i = (c_1\odot a_{i-1}) \oplus (c_2 \odot a_{i-2}) \oplus ... \oplus (c_{n} \odot a_{i-n}) = \sum_{k=1}^{n}c_{k}a_{i-k}.$$ 
In the above equation, all the variables may assume only binary values (e.g., 1 or 0), and $\odot$ and $\oplus$ are {\em modulo-2} multiplication and {\em modulo-2} addition operations, respectively. Specifically, the generated m-sequences with non-zero initial vectors (i.e., {\em seeds}) have period $N = 2^n - 1$ with the following three randomness properties that minimize the interference.
\\
\indent {\bf (1)} The number of 1's and 0's are approximately equal. 
\\
\indent {\bf (2)} Half of the runs (i.e., subsequences of consecutive 1's and consecutive 0's) have length 1, $\frac{1}{4}$ runs have length 2, $\frac{1}{8}$ runs have length 3, and $\frac{1}{2^k}$ have length $k$, where $(k < n)$. 
\\
\indent {\bf (3)} It has sequence {\em autocorrelation} that is a randomness measurement and provides the degree of correspondence between an m-sequence and its phase-shifted replica. The smaller the correlation, the easier it is for a receiver to recover the m-sequence from interference.

The periodic autocorrelation function \(R\) of an m-sequence is given by 
$$R(\tau) = \frac{1}{N}\sum_{n=1}^{N}a^{'}_{n}a^{'}_{n-\tau}$$ 
where $a^{'}_n = 1-2a_n$ (i.e., a $\pm1$ sequence) and $\tau$ represents different periods. It can also be shown that the periodic autocorrelation of an m-sequence is
  \begin{equation}
  	R(\tau) = \begin{cases}
    1 &\text{$\tau = 0, N, 2N,...$}\\
    -\frac{1}N &\text{otherwise.} 
    \end{cases}\nonumber 
  \end{equation}

Similar to autocorrelation, cross-correlation is also the measurement of interest in m-sequences. It is the degree of correspondence between m-sequences used by different users (i.e., sensors). Intuitively, the cross-correlation between different m-sequences needs to be low to avoid interference. If $a^{'}_n$ and $b^{'}_n$ are two m-sequences, then their cross-correlation
\vspace{-0.1in}
\begin{equation}
R_{a^{'},~b^{'}}(\tau) = \frac{1}N \sum_{n=1}^{N}a^{'}_{n}b^{'}_{n-\tau} \nonumber
\label{eqn:cross}
\end{equation}
where $b^{'}_n = 1-2b_n$ (i.e., a $\pm 1$ sequence). \revise{It has been shown that the number of m-sequences that have the least cross-correlation values between themselves is very small and may not be feasible to be used in multiple access systems~\cite{sarwate1980crosscorrelation, lee2000sequence}, including D-OFDM in SNOW due to the {\em asynchronicity} between the sensors within and across the subcarriers. To this extent, we generate a set of PN sequences based on Gold code~\cite{dinan1998spreading} using the generated m-sequences above.}

\subsection{Gold Code-Based PN Sequences in mSNOW}
Similar to the {\em Gold codes} in DS-CDMA (direct sequence code division multiple access), we generate a set of PN sequences for the D-OFDM system in mSNOW such that different sensors may transmit or receive {\em asynchronously} and {\em concurrently} within and across the subcarriers (which is very much unlike DS-CDMA). Gold codes provide a uniform and bounded cross-correlation between the codes~\cite{fan1996sequence, pradhan2014novel, gold1968maximal}. Similar to the Gold codes, our PN sequences are generated by repeatedly taking bitwise XOR of two {\em uncorrelated} m-sequences of the same length, which are generated by two LFSRs (say, \code{LFSR}$_1$ and \code{LFSR}$_2$), respectively. 
\begin{figure}[!htbp]
    \centering
    \includegraphics[width=0.45\textwidth]{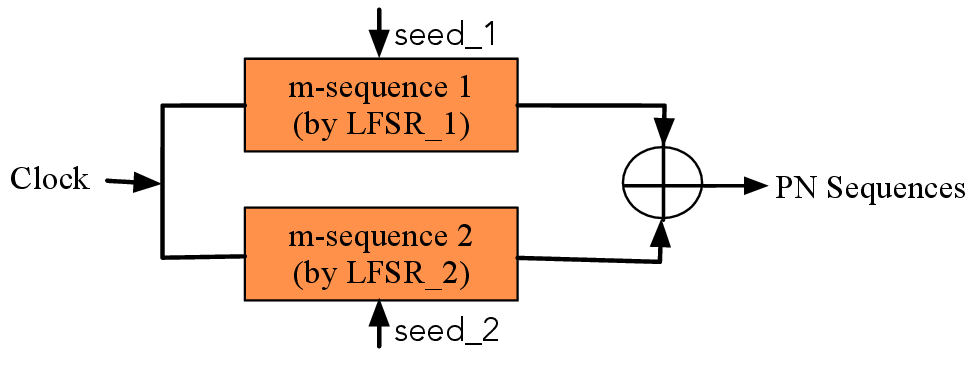} 
    \caption{Generation of PN sequences in mSNOW.}
    \label{fig:gcgenerator}
\end{figure}

\revise{Figure~\ref{fig:gcgenerator} shows such a generator, where \code{LFSR}$_1$ and \code{LFSR}$_2$ use two non-zero seeds \code{seed}$_1$ and \code{seed}$_2$ (each of length $n$), respectively. Note that an LFSR with a particular seed of length $n$ bits generates an m-sequence of length $N = 2^n -1$ bits. 
Consequently, the length of a PN sequence generated by two uncorrelated m-sequences from two different LFSRs (as shown in Figure~\ref{fig:gcgenerator}) is also $N = 2^n -1$ bits.
Figure~\ref{fig:gcgenerator} also confirms that changing the seeds of the LFSRs generates new sets of PN sequences. For each PN sequence in a set, there may exist many pairs of m-sequences. However, not each pair of m-sequences generates a PN sequence that may have the least cross-correlation values (i.e., less mutual interference) with the other PN sequences in the same set. For this, the PN sequences in D-OFDM should have three-valued peak cross-correlation magnitudes that are both uniform and bounded~\cite{sarwate1980crosscorrelation}.}

To generate a set of PN sequences with the above requirements, a good pair of m-sequences (a.k.a. {\em preferred pair}) is needed. Let our preferred pair be $\{u,v\}$ where $u$ and $v$ are generated by \code{LFSR}$_1$ and \code{LFSR}$_2$, respectively. If we consider $u$ as a binary vector, then $v$ can be generated in a deterministic manner by sampling every $q$-th bit of $u$, for some appropriate $q$ (e.g., if and only if $gcd(N, q) = 1$~\cite{dinan1998spreading}) from multiple copies of $u$ until both $u$ and $v$ have the same length $N=2^n -1$ bits. The $i$-th PN sequence is then generated by a bitwise XOR of $u$ and an $i$-bit shifted copy of $v$. Specifically, $\{u,v\}$ should have the following properties.
\\
\indent {\bf (1)} Both \code{LFSR}$_1$ and \code{LFSR}$_2$ have preferred but unique polynomial equations with a degree of $n$ (i.e., equal to the length of their seeds).
\\
\indent {\bf (2)} $n$ is not divisible by 4~\cite{golomb1994shift}. \revise{When $n$ is a multiple of 4, it becomes significantly more difficult, or even impossible, to find a pair of polynomial equations (i.e., a preferred pair of m-sequences) that result in PN sequences with the desired low cross-correlation properties.}
\\
\indent {\bf (3)} $q$ is odd and either $q = (2^k+1)$ or $q = (2^{2k}–2^k+1$). 
\\
\indent {\bf (4)} $gcd(n, k) = 1$ if $n$ is odd or $gcd(n, k) = 2$ is $n$ is even.

\noindent Using the above technique, the set of PN sequences generated may be denoted as follows: $$G(u, v) = \{u, v, u \oplus v, u \oplus Dv, u \oplus D^2v,..., u \oplus D^{N-1}v\}.$$ In the above equation, $D$ is the delay element and represents the operator that shifts vectors cyclically to the left by one place. Additionally, $G(u,v)$ contains a total of $M = (N + 2)$ PN sequences, where $N = (2^n - 1)$ and the "$+2$" term is for the initial preferred pairs. In $G(u,v)$, any pair of PN sequences or a PN sequence and its shifted version has one of the three cross-correlation magnitudes in $\{–t(n)$, $–1$, $t(n) – 2\}$, where 
\[t(n) = \begin{cases}
    1+2^{(n+2)/2} &\text{$n$ even}\\
    1+2^{(n+1)/2} &\text{$n$ odd.}
    \end{cases} \]

\revise{In the following, we now present our techniques for encoding and decoding data in both uplink and downlink communications within the mSNOW framework. Specifically, we first generalize the signal-level encoding and decoding processes and then attribute these techniques to both the uplink and downlink communications in mSNOW, as appropriate, based on our system model.}

\subsection{Encoding the Transmitted Signal}
As discussed in Section~\ref{sec:model1}, we consider ASK, especially OOK (on-off keying), as the  D-OFDM subcarrier modulation technique in mSNOW, where presence and absence of a carrier signal represent bit 1 and bit 0, respectively. Within a D-OFDM subcarrier, a sensor thus transmits a signal (which is termed a {\em symbol} in D-OFDM) or refrains from it to represent a data bit 1 or 0, respectively. Typically in SNOW, a data bit is spread to 8 bits (which constitute the actual symbol duration) by repeating it 8 times to strengthen (e.g., along with subcarrier orthogonality) the resistance against inter-subcarrier interference by creating an effect similar to the {\em cyclic-prefix}-based guard bands used in single-user OFDM systems~\cite{saifullah2018low}.
In our design, we spread a data bit to $N$ bits by repeating it $N$ times and then multiplying the sensor's PN sequence (and subsequently mixed with the subcarrier signal). In mSNOW, we thus have an $N$-bit long symbol that accounts for both inter-symbol interference (between the sensors within a subcarrier) and inter-subcarrier interference (between sensors across subcarriers). This enables our proposed concurrency in the SNOW PHY layer, as shown in Figure~\ref{fig:model}, where multiple asynchronous sensors is able to concurrently transmit or receive within and across the D-OFDM subcarriers.

To this extent, let $b_{ij}(k)$ and $g_{ij}(k)$ be the $k$-th spread-bit of a data bit and the $k$-th bit of the PN sequence of {\em $j$-th sensor on $i$-th subcarrier}, respectively. Thus, the signal for the $k$-th spread-bit is $x_{ij}(k) = b_{ij}(k)g_{ij}(k)$. Overall, the symbol for a data bit 1 in our design may be represented as 
\begin{equation}
[g_{ij}(k), g_{ij}(k+1),..., g_{ij}(k+N-1)]^T = g_{ij}.
\label{eqn:symbol-encoding}
\end{equation}
We can create an equal-length (i.e., $N$-bit) symbol for data bit 0 with the similar process, which will be all 0's, and hence no signal transmission when mixed with the subcarrier signal.

\subsubsection{Aggregate Signal in Uplink Communication}
\revise{With the symbol-level signal representation in Equation (\ref{eqn:symbol-encoding}), we may now construct the aggregate transmitted signal from all the asynchronous sensors during the uplink communication. Let each spread-bit within each symbol of a packet containing $l$ total symbols transmitted by the $j$-th sensor on the $i$-th subcarrier be represented by exactly one discrete-time sample. A packet by the $j$-th sensor on the $i$-th subcarrier may thus be ideally (e.g., without any noise contributions) represented, using the time-shifting property of signal, as 
\begin{equation}
p_{ij}[k] = \sum_{m=1}^{l}g_{ij}[k-m].
\label{eqn:packet-encoding}
\end{equation}
Consequently, the aggregate transmitted signal from $L$ number of asynchronous sensors concurrently transmitting packets on the $i$-th subcarrier may be ideally represented as 
\begin{equation}
p_i[k] = \sum_{j=1}^{L} p_{ij}[k].
\label{eqn:subcarrier-encoding}
\end{equation}
Finally, the aggregate signal at the BS from all the D-OFDM subcarriers, assuming $L$ asynchronous sensors concurrently transmitting on each subcarrier $i$ with center frequency $f_i$, may be represented as follows:
\begin{equation}
\sum_{i=1}^{M} p_{i}[k]~e^{\sqrt{-1}.2 \pi f_i t} + Z_M
\label{eqn:dofdm-encoding}
\end{equation}
where $M$ is the total number of orthogonal subcarriers in mSNOW and $Z_M$ is the additive white Gaussian noise vector (AWGN) for all the subcarriers. To avoid confusions, we write $\sqrt{-1}$ to represent the unity of imaginary numbers since both the letters $i$ and $j$ have been used to denote other aspects in the equations. Additionally, if needed, Equation (\ref{eqn:dofdm-encoding}) may be generalized for any number of packets by any sensor on any subcarrier.}

\subsubsection{Aggregate Signal in Downlink Communication}
\revise{In downlink communication, the SNOW BS transmits an aggregate signal using its Tx-Radio (as shown in Figure~\ref{fig:model}) containing distinct data (if any) for different sensors listening asynchronously on different D-OFDM subcarriers. For this, the BS creates an OFDM signal by applying IFFT on the available data for the intended sensors. We use a similar technique as used in SNOW for creating the aggregate signal in downlink communication in mSNOW; however, note that the BS should be able to encode data for multiple sensors listening within the same subcarrier  as well (which is unlike SNOW). This may be done easily in mSNOW BS by reusing the steps described in Equations (\ref{eqn:symbol-encoding})--(\ref{eqn:subcarrier-encoding}). Finally, the composite signal for different sensors listening within and across the D-OFDM subcarriers may be represented using the following time-domain representation:
\begin{equation}
\frac{1}{\sqrt{M}}\sum_{i=0}^{M-1}p_i~e^{\sqrt{-1}.2 \pi f_i t}
\label{eqn:bs-encoding}
\end{equation} 
where $M$ is the number of subcarriers, $p_i$ is formed using the steps shown in Equation (\ref{eqn:subcarrier-encoding}), $f_i$ is the center frequency of the $i$-th subcarrier, and time $t$ accounts for the (composite) packet durations across all the D-OFDM subcarriers in mSNOW.}

\subsection{Decoding the Received Signal}
\subsubsection {Decoding at the BS in Uplink Communication} As the asynchronous sensors may concurrently transmit within and across the D-OFDM subcarriers in mSNOW, Equation (\ref{eqn:dofdm-encoding}) represents the received signal at the BS Rx-Radio. To decode at the subcarrier level, the BS applies a global FFT algorithm (i.e., G-FFT) on the received signal, similar to the technique in SNOW.
After the G-FFT, samples in each subcarrier are isolated (from the corresponding FFT bins) and considered for despreading and decoding (which is unlike SNOW) in our design. Let $r_i$ be the received samples' vector of a symbol at the $i$-th subcarrier after G-FFT. Each sample $k$ in $r_i$ may be represented as
\begin{equation}
r_i[k] =  \sum_{j=1}^{L}x_{ij}[k] + z[k] = \sum_{j=1}^{L}b_{ij}g_{ij}[k] + z[k]
\label{eqn:fft-bin}
\end{equation}
where $L$ is the number of sensors using subcarrier $i$ and $z$ is the additive white Gaussian noise vector. Note that the power level (i.e., magnitude) of each sample is given by the G-FFT algorithm. Similar to the current SNOW PHY demodulator~\cite{saifullah2016snow}, we maintain a 2D matrix at the BS to decode each data bit from each sensor in each subcarrier in an asynchronous fashion. An entry $r_i(k)$ (interpreted as $r[i][k]$) in the matrix represents the $k$-th sample in $i$-th subcarrier. A decoding agent in the BS keeps running in the background to detect, decode, and despread data from different sensors on each subcarrier by multiplying different PN sequences for that subcarrier. For example, the despread data from the $j$-th sensor  on the $i$-th subcarrier may be represented as
\begin{equation}\footnotesize
r_i^Tg_{ij} = [r_i(k), r_i(k+1),..., r_i(k+N-1)]\begin{bmatrix}
g_{ij}(k)\\
.\\
.\\
.\\
g_{ij}(k+N-1)
\end{bmatrix}.
\label{eqn:final-decoding}
\end{equation}
The above operation gets rid of the interference by the other sensors (if any) sharing the $i$-th subcarrier along with any other noise.
Note that the vectors of samples of symbols are generated right after the detection of a preamble of the packets in the subcarriers. After a symbol is despread, we recover the original data bit (which was repeated before spreading) by simply undoing the repeat operation. For this, we consider that a data bit is 1 if at least half of the repeated bits remain 1. This technique allows for an additional guard against interference.

\subsubsection {Decoding at the Sensors in Downlink Communication}
\revise{In downlink, the BS may transmit to multiple asynchronous sensors listening concurrently to a subcarrier as well as to other sensors listening to the other subcarriers in a similar fashion. For this, the BS makes a single transmission of a composite signal that spans (i.e., its bandwidth) across the frequencies of all the subcarriers, which may be represented by Equation (\ref{eqn:bs-encoding}). An asynchronous sensor, however, receives only the portion of the composite signal, especially the part that was encoded on its subcarrier's center frequency by the BS. As expected, multiple sensors listening to the same subcarrier receive the same portion of the composite signal. They, however, may despread and decode their data independently and asynchronously using the steps encompassing Equation (\ref{eqn:fft-bin}) and then Equation (\ref{eqn:final-decoding}). Note that a sensor decoding its data {\em need not} employ the FFT algorithm before applying Equations (\ref{eqn:fft-bin}) and (\ref{eqn:final-decoding}).}

\subsection{Analyzing Achievable Bitrate in mSNOW}\label{sec:ourbitrate}
In this section, we theoretically analyze the bitrate in mSNOW.
For a sensor on an AWGN subcarrier of bandwidth $B$ with signal-to-noise ratio (SNR) $SNR$, the maximum Tx bitrate $C = B\log_2(1 + SNR)$ based on the Shannon-Hartley Theorem~\cite{shannon2006communication}. On a subcarrier with $B = 200$ kHz and $SNR = 3$ dB, we may achieve a bitrate of $\frac{200\times2}{N} = \frac{400}{N}$ kbps (recall that $N$ is the PN sequence length of the sensor). In our evaluations (Section~\ref{sec:eval}), we choose $N = 7$, which gives us a Tx bitrate of $\approx$57.14 kbps per sensor. Additionally, two signal levels in our ASK modulation conform to the Nyquist Theorem $C = 2B\log_22^m$ where $2^m$ is the number of signal levels to support a theoretical bitrate of $\approx$57.14 kbps per sensor. If a subcarrier is shared by $M = (N + 2)$ number of sensors, then our maximum achievable Tx bitrate over bandwidth $B$ increases $M$-times, which is $\approx$M-times better (and conforms to the Nyquist Theorem) compared to the IEEE 802.15.4 standards' datarate requirements of 50 kbps over a channel~\cite{ieee802154} or the current SNOW design. In evaluations, we, however, choose $B = 400$ kHz due to interference created by concurrent Txs and an \code{SNR} = 6 at the BS, which still provides us with an effective bitrate of $>$ 50 kbps per sensor.

\subsection{Analyzing Scalability in mSNOW}\label{sec:ourscalability}
\revise{As discussed in Section~\ref{sec:related_lora}, LoRa or Sigfox achieves scalability assuming very low traffic. An SX1301 LoRa gateway having 8 in-built radios for concurrent transmissions on 8 channels may have 62,500 sensors, given that each sensor transmits one packet every hour~\cite{saifullah2016snow}. A Sigfox gateway, on the other hand, may support 1 million sensors if each sensor transmits 140 12-byte packets per day~\cite{sigfox}. In mSNOW, having a single TV white space channel (i.e., having 6 MHz bandwidth) split into twenty-nine 400 kHz D-OFDM subcarriers allows $(9\times29) = 261$ sensors to transmit {\em concurrently} (assuming a PN sequence of length 7) in less than 2 ms (considering 12-byte packets). Thus, in uplink communication in mSNOW, a single D-OFDM subcarrier may be shared by at least $\frac{9\times24\times3600\times1000}{140\times2} \approx 2,777,142$ sensors if each sensor transmits only 140 packets per day (as in Sigfox), totaling approximately $(29\times2,777,142 = 80,537,118$ or $80.5$ million sensors on 29 D-OFDM subcarriers. Considering both uplink and downlink communications together, mSNOW may still support approximately $(80.5 / 2) = 40.25$ million sensors if a set of uplink transmissions by a group of sensors is immediately followed by a set of downlink transmissions to the same group of sensors. If mSNOW can acquire m TV white space channels, then it may support approximately $(40.25\times m)$ million sensors. This back-of-the-envelope calculation thus suggests a significantly higher scalability in mSNOW compared to both LoRa and Sigfox. Finally, mSNOW may be at least 9x more scalable than SNOW in the above scenarios.}

\section{Evaluation}\label{sec:eval}
In this section, we present our implementation and evaluate our design for various link parameters and network parameters.

\subsection{Implementation Platform}
We create a mSNOW simulation platform using the Python programming language. For splitting a wideband into narrowband AWGN subcarriers, performing FFTor IFFT operations, and other signal processing operations, we use the Python NumPy library. Additionally, we use the Python NumPy library for generating the PN sequences based on our design and encoding and decoding processes at the sensors and the BS for both the uplink and downlink communications in mSNOW.  Our open-source implementation is available online~\cite{snowsimulation}.

\subsection{Evaluation Setup and Default Parameter Settings}\label{sec:eval_setup}
As discussed in our system model in Section~\ref{sec:model1}, no two sensors sharing the same subcarrier or are on neighboring subcarriers do not have the same PN sequence assigned to them. To ensure this, we generate two instances of our set of PN sequences (say, $PNs_1$ and $PNs_2$) using two different sets of initial seeds, which still hold the required cross-correlation properties. We then allocate $PNs_1$ to all the odd-numbered subcarriers and $PNs_2$ to all the even-numbered subcarriers, and thus ensuring the above requirements. \revise{For both $PNs_1$ and $PNs_2$, we use $n=3$, which yield $N=7$, and thus aim for the discussed datarate in Section~\ref{sec:ourbitrate}. For $PNs_1$, we choose \code{seed}$_1$ = \code{seed}$_2$ = 101 (both can be the same since LFSRs use different polynomial equations) and get $PNs_1 = $
\{1011100, 1010011, 0001111, 1111011, 0010010, 1000001, 1100110, 0101000, 0110101\}. For $PNs_2$, we choose \code{seed}$_1$ = \code{seed}$_2$ = 010 (which is different from the seeds of $PNs_1$) and get $PNs_2 = $ \{0101110, 0100111, 0001001, 1100000, 0110011, 0010100, 1011010, 1000111, 1111101\}.} 

\revise{In our evaluation, we use sixty-four 400 kHz subcarriers (numbered 1 -- 64) with 50\% overlapping within 547 MHz -- 560 MHz (i.e., a chunk of the TV white space spectrum), and each subcarrier is shared by at most 9 sensors, totaling 576 sensors (which is $9$-times higher than the original SNOW could accommodate in its PHY layer). In general, the equation $\frac{W}{\omega \alpha} - 1$, where $W$ is the used bandwidth of the TV white space spectrum, $\omega$ is the subcarrier bandwidth, and $\alpha$ is the subcarrier carrier overlapping factor (which may be a maximum of 50\% to maintain subcarrier orthogonality in D-OFDM), gives us the total number of usable subcarrier in mSNOW.
The 64 subcarriers in our setup thus have center frequencies 547.2 MHz, 547.4 MHz, 547.6 MHz, $...$, 559.4 MHz, 559.6 MHz, and 559.8 MHz, respectively.
Similar to the current SNOW, we emulate a Tx power of 0 dBm, receive sensitivity of -85 dBm, packet size of 40 bytes (excluding an 1-byte preamble) containing 12-byte header and 28-byte random payload (data, CRC), OOK (on-off keying) for subcarrier data modulation, and an SNR of 6 dB. Unless stated otherwise, these are our default parameter settings.}

\subsection{Threshold Selection}
\revise{In our evaluation, we first decide on the threshold values that may be used in different subcarriers in mSNOW to denote the presence or absence of data, which is crucial for our design.
Since at most 9 sensors may transmit on the same subcarrier in mSNOW, the received signal strength (RSS) of the received symbols after the FFT output is not limited to 0s and 1s (as in SNOW)}. Signals from concurrently transmitting sensors {\em superimpose} and make it challenging to decide the magnitude of the composite signal. For this, we consider the average signal power $\sum_{i=1}^{M}\sqrt{I^2 + Q^2}$ to decide on different thresholds levels, where $I$ and $Q$ are the in-phase and quadrature signal components, and $M$ is the averaging number of samples. Specifically, we collect 50,000 samples for each case when the number of concurrently transmitting sensors on a  subcarrier vary from 0 to 9, where each sensor also transmits only 1s. 

\begin{figure}[!htbp]
    \centering 
    \includegraphics[width=0.4\textwidth]{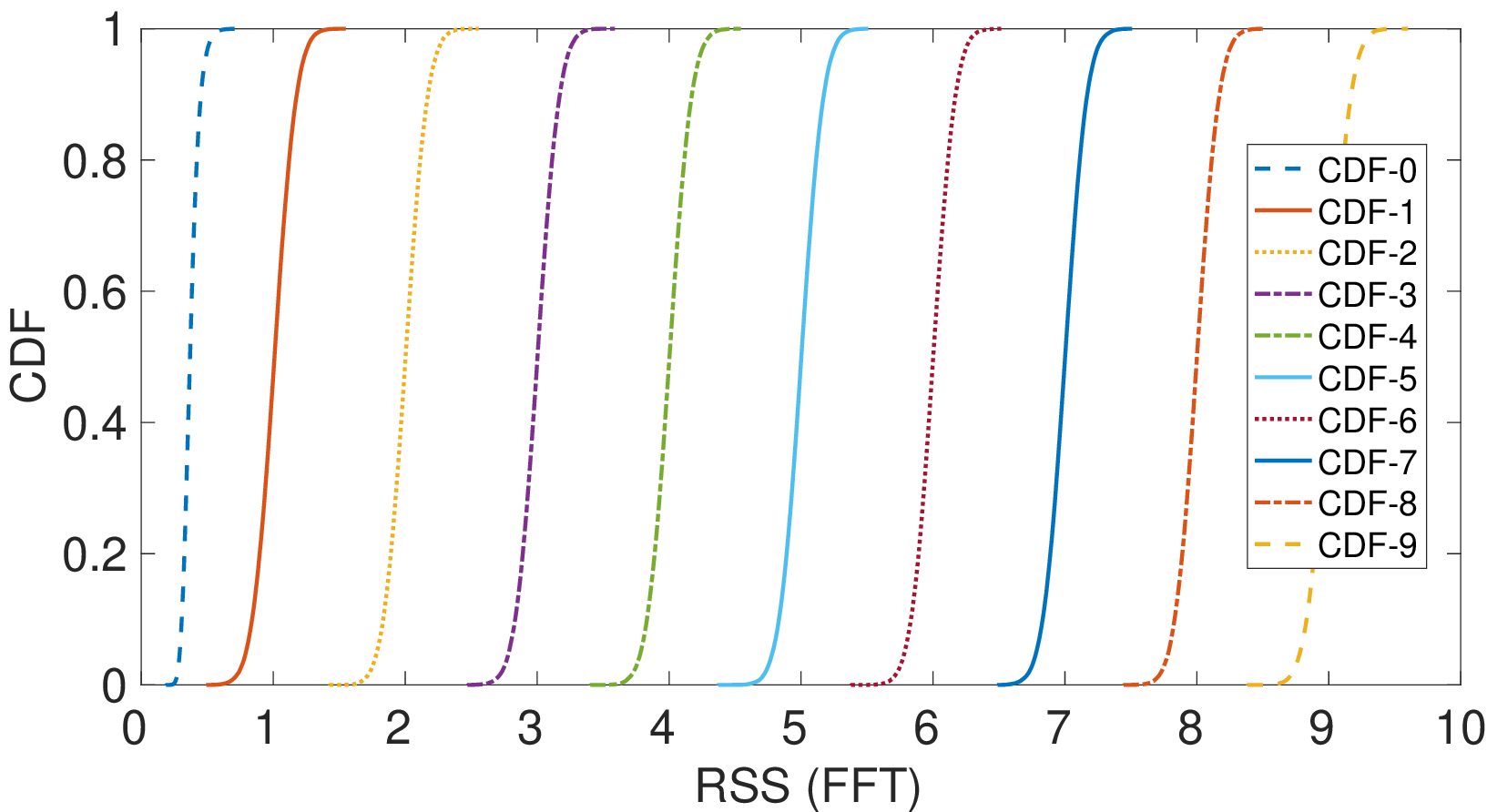}
    \caption{Threshold behavior in mSNOW.}
    \label{fig:threshold}
\end{figure}
Figure~\ref{fig:threshold} shows the {\em cumulative distribution function} (CDF) of RSS values in the above setup. As shown in this figure, when there is no transmission on a subcarrier, the RSS is below 0.5 for 100\% of the cases, which may be used to denote the 0 magnitude or the noise floor in our evaluation. When a single sensor transmits, the RSS is between 0.51 and 1.5 for 100\% of the cases, which may be quantized to magnitude 1. For the case of 2 sensors, the RSS is between 1.51 and 2.5 for 100\% of the cases, which may be quantized to magnitude 2. Similarly, to denote magnitudes 3, 4, 5, 6, 7, 8, and 9, the RSS ranges are 2.51 -- 3.5, 3.51 -- 4.5, 4.51 -- 5.5, 5.51 -- 6.5, 6.51 -- 7.5, 7.51 -- 8.5, and 8.51 -- 9.5 for 100\% of the cases, respectively. \revise{A similar behavior of the RSSI in the downlink communication (e.g., while the BS is transmitting a composite signal on a subcarrier for different numbers of sensors) may be observed as well in mSNOW.
In the rest of the evaluations, we use the findings in this section to determine different magnitude levels, as necessary for despreading and decoding in both uplink and downlink communications in mSNOW.}

\subsection{Evaluating Link Performance}\label{sec:linkperformance}
\revise{In this section, we evaluate the D-OFDM subcarrier link reliability in both uplink and downlink communications in mSNOW.} For this evaluation, we consider 3 neighboring subcarriers with 50\% overlaps, which may generalize all the subcarriers in our implementation. For example, we use subcarriers with center frequencies 549.8 MHz, 550.0 MHz, and 550.2 MHz, where the subcarrier with 550.0 MHz center frequency is the middle subcarrier and overlaps 50\% with its neighbors on both sides. 
\revise{As a reliability metric, we use {\em correctly decoding rate} (CDR), which refers to the percentage of packets that are correctly decoded at the BS (in uplink communication) among all the transmitted ones by the sensors or the percentage of packets that are correctly decoded at a sensor (in downlink communication) among all the transmitted ones by the BS.}

\subsubsection{Link Reliability in Uplink Communication}
In the setup for uplink communication, we allow 1 to 9 sensors concurrently transmit on each of the considered subcarriers, totaling 27 sensors. In each case, a node sends consecutive 100 40-byte packets (with random payloads) to the BS using its subcarrier with a random inter-packet interval of 0 -- 3 ms that ensures overlapping of packets (each packet takes $\approx$5.6 ms to transmit) with other sensors on the same or neighboring subcarriers. We also repeat this experiment 100 times.

\begin{figure}[!htbp]
    \centering
      \subfigure[Average correctly decoding rate (uplink)\label{fig:avg_crds}]{
    \includegraphics[width=0.3\textwidth]{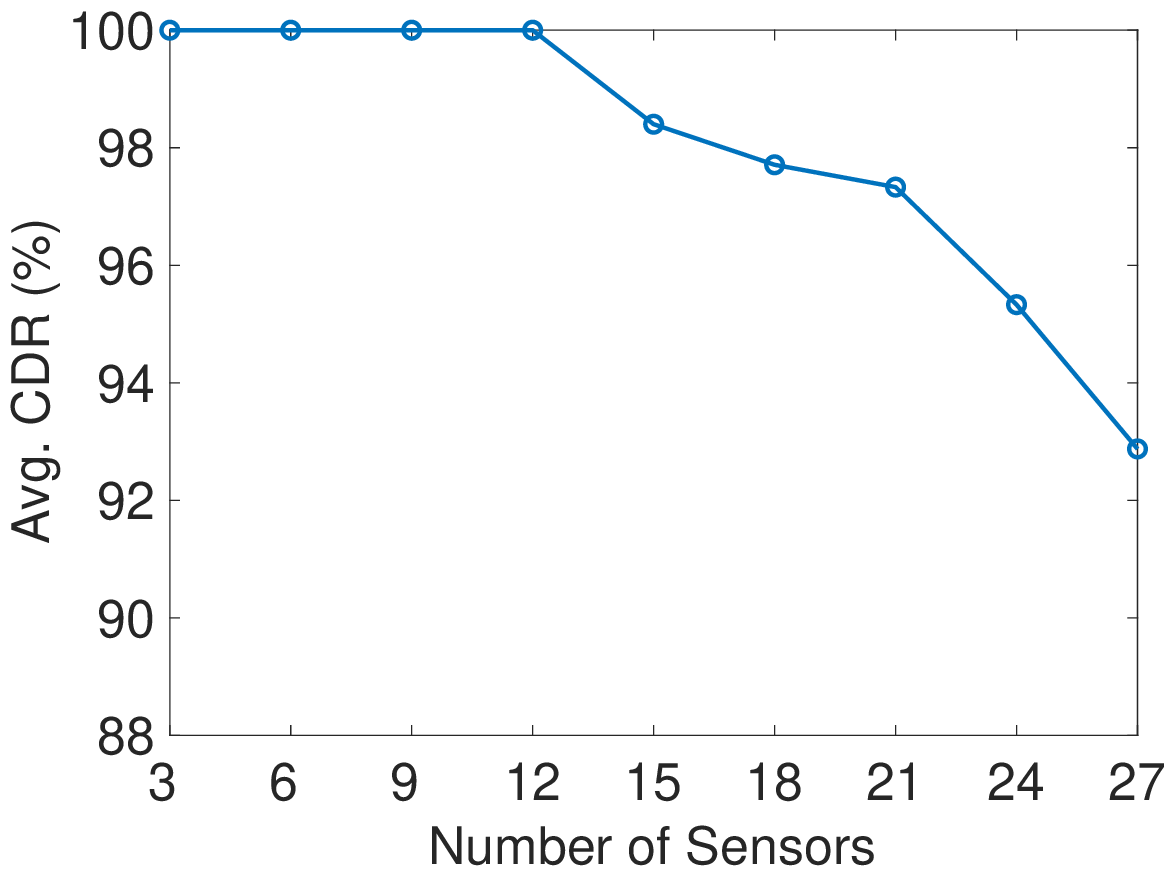}
     }\vfill 
      \subfigure[CDR in subcarriers (extreme case in uplink)\label{fig:cdrs}]{
        \includegraphics[width=.3\textwidth]{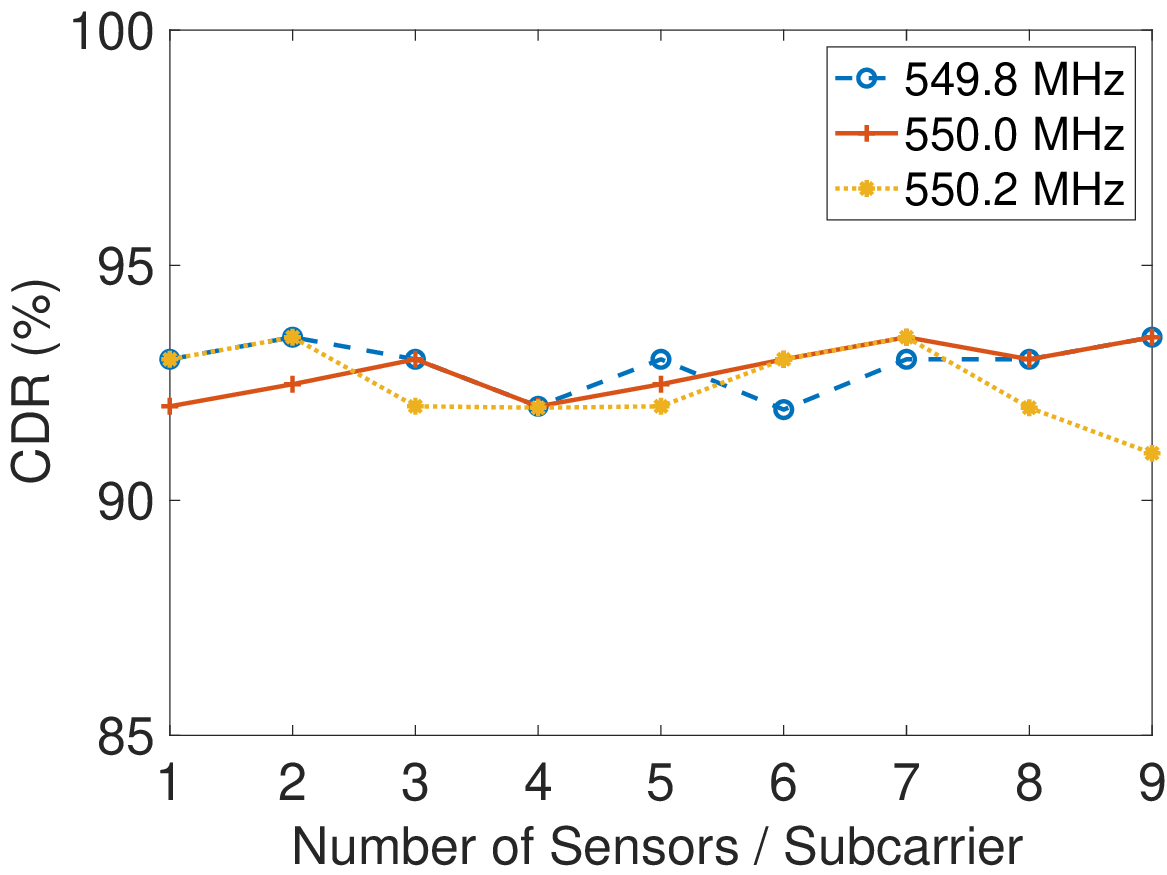}
      }
    \caption{Link reliability in uplink communications in mSNOW.} 
    \label{fig:linkperform}
\end{figure}

\noindent{\bf Correctly Decoding Rate in Uplink.}
Figure~\ref{fig:avg_crds} shows the average CDR at the BS for the selected subcarriers for different number of sensors on different subcarriers. As shown in this figure, when 3, 6, 9, and 12 sensors transmit (i.e., 1, 2, 3, and 4 sensors on each subcarrier, respectively), the average (across all the subcarriers) CDRs of all these cases are approximately 100\%. For the cases where 15, 18, 21, 24, and 27 sensors transmit (i.e., 5, 6, 7, 8, and 9 on each subcarrier, respectively), the average CDRs are approximately 98.4\%, 97.71\%, 97.33\%, 95.33\%, and 92.88\%, respectively. In summary, in all the cases, the average CDRs are above 92\%, which confirm high reliability of our design under massive concurrency and is acceptable in  wireless networks~\cite{rahman2021lpwan, saifullah2018low}.

\noindent{\bf Subcarrier Reliability in Extreme Case in Uplink.}
Figure~\ref{fig:cdrs} shows the CDRs on different subcarriers for the extreme case where 27 sensors transmit concurrently (i.e., 9 sensors on each subcarrier) and there are no inter-packet (of different senors) delays. Thus, all the packets are colliding in the worst way possible in a network with concurrent transmissions in this scenario. As shown in Figure~\ref{fig:cdrs}, for the packets (approximately 30,000 40-byte packets with random payloads) of 1 sensor on each subcarrier (i.e., total 3 sensors), the CDRs on 549.8 MHz subcarrier is approximately 93\%, 550.0 MHz subcarrier is approximately 92\%, and 550.2 MHz subcarrier is approximately 93\%. This figure also shows that, as we increase the number of sensors on each subcarrier, the CDRs do not change drastically. For example, the extreme cases of 2, 3, 4, 5, 6, 7, 8, and 9 sensors on each subcarrier also yield CDRs in the approximate range of 92\% -- 93\% in all the three selected subcarriers in this setup.

\begin{figure}[!htbp]
    \centering 
    \includegraphics[width=0.3\textwidth]{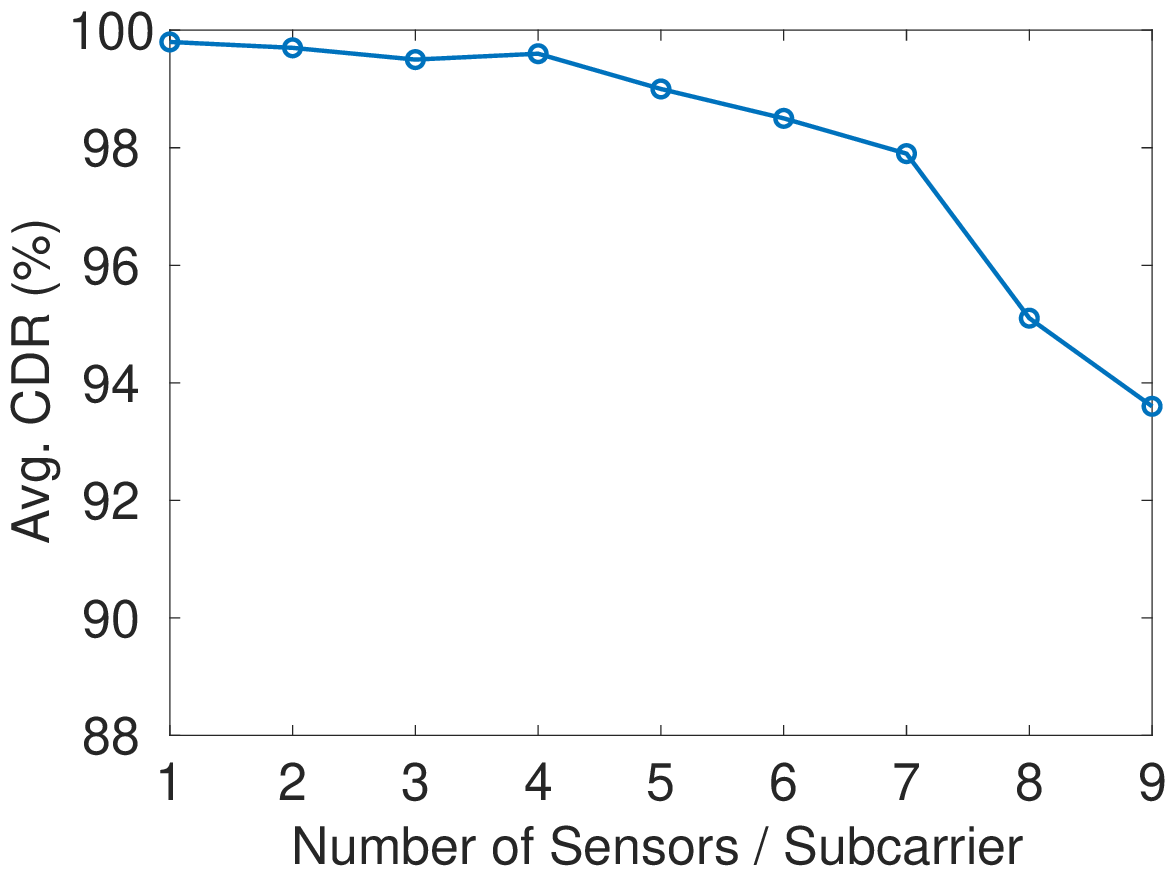}
    \caption{Link reliability in downlink communications in mSNOW.}
    \label{fig:dlink_cdrs}
\end{figure}
\subsubsection{Link Reliability in Downlink Communication}
\revise{To evaluate the downlink communications in mSNOW, we allow the BS to concurrently transmit to 1--9 sensors on each of the three subcarriers. Specifically, the BS sends consecutive 100 40-byte packets (with random payloads) to the sensors on their subcarriers with a random inter-packet interval of 0 -- 3 ms that ensures overlapping of packets within and across the selected subcarriers. Figure~\ref{fig:dlink_cdrs} shows the average (in the sensors across all the subcarriers) CDRs in the sensors on for varying numbers of sensors with the above transmission pattern repeated 100 times. 
As shown in this figure, when the BS concurrently transmits to 1, 2, 3, and 4 sensors on each subcarrier, the average CDRs at the sensors are close to 100\%. For the cases where the BS concurrently transmits to 5, 6, 7, 8, and 9 on each subcarrier, the average CDRs are approximately 99\%, 98.49\%, 97.9\%, 95.12\%, and 93.61\%, respectively. In summary, the average CDRs consistently exceed 93\% in all cases, indicating the high reliability of mSNOW in concurrent downlink communications.}

\begin{figure*}[!htbp]
    \centering
      \subfigure[Overall network throughput\label{fig:nthroughput}]{
    \includegraphics[width=0.315\textwidth]{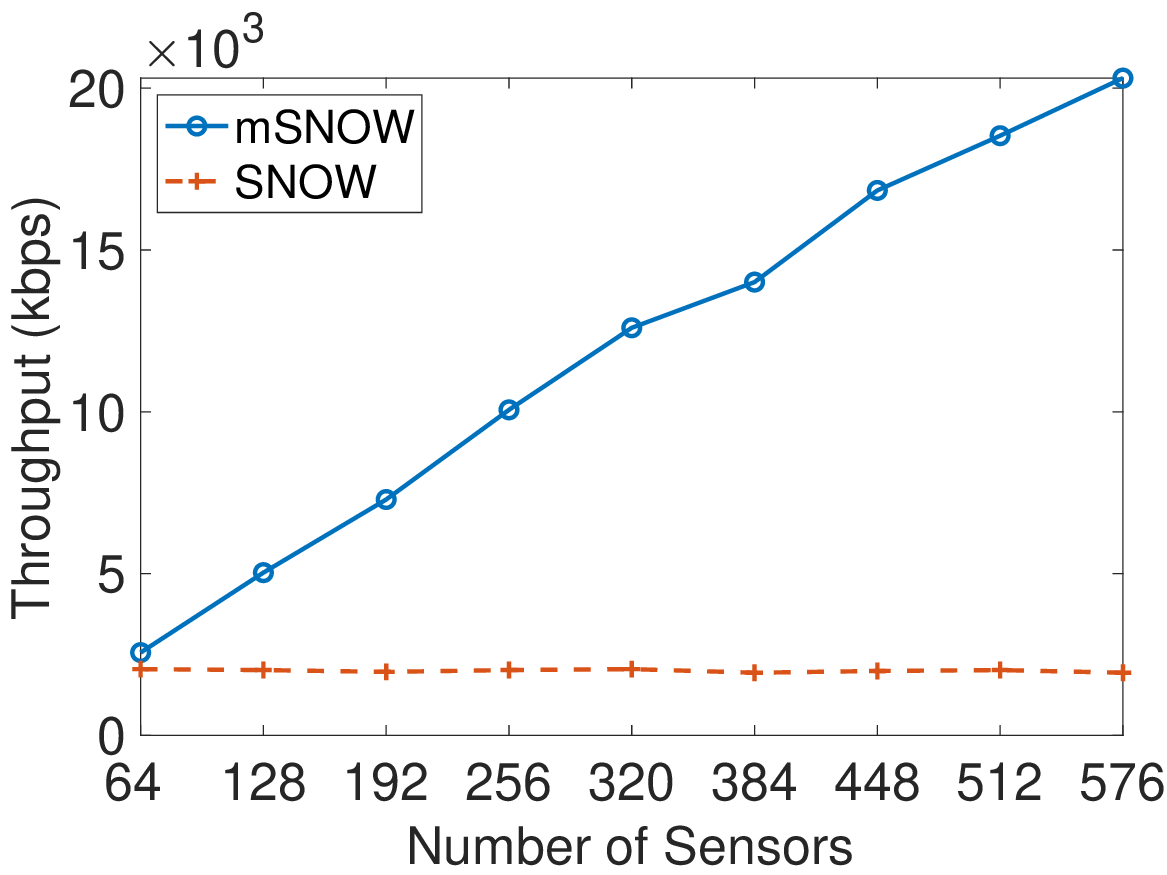}
      }\hfill 
      \subfigure[Average latency per packet\label{fig:nlatency}]{
        \includegraphics[width=.3\textwidth]{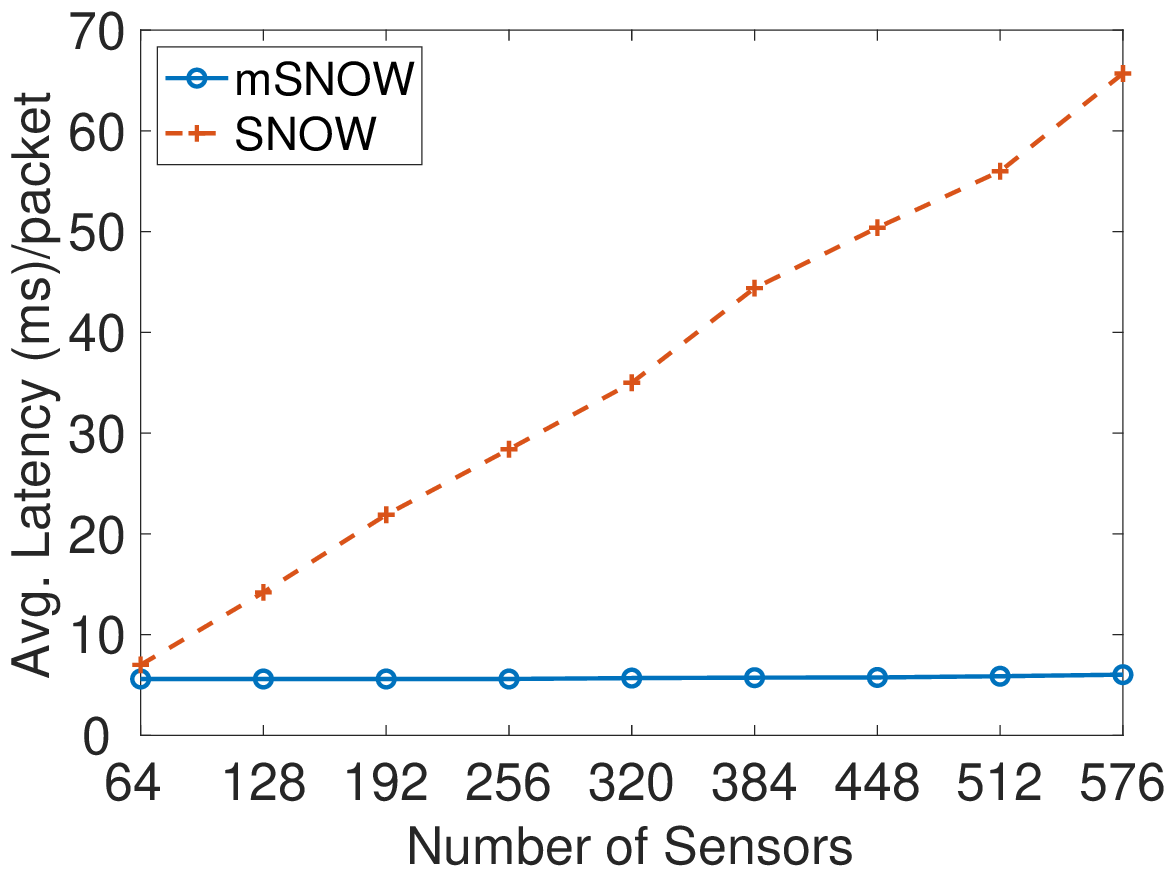}
      }
      \hfill 
      \subfigure[Average energy consumption per packet\label{fig:nenergy}]{
        \includegraphics[width=.3\textwidth]{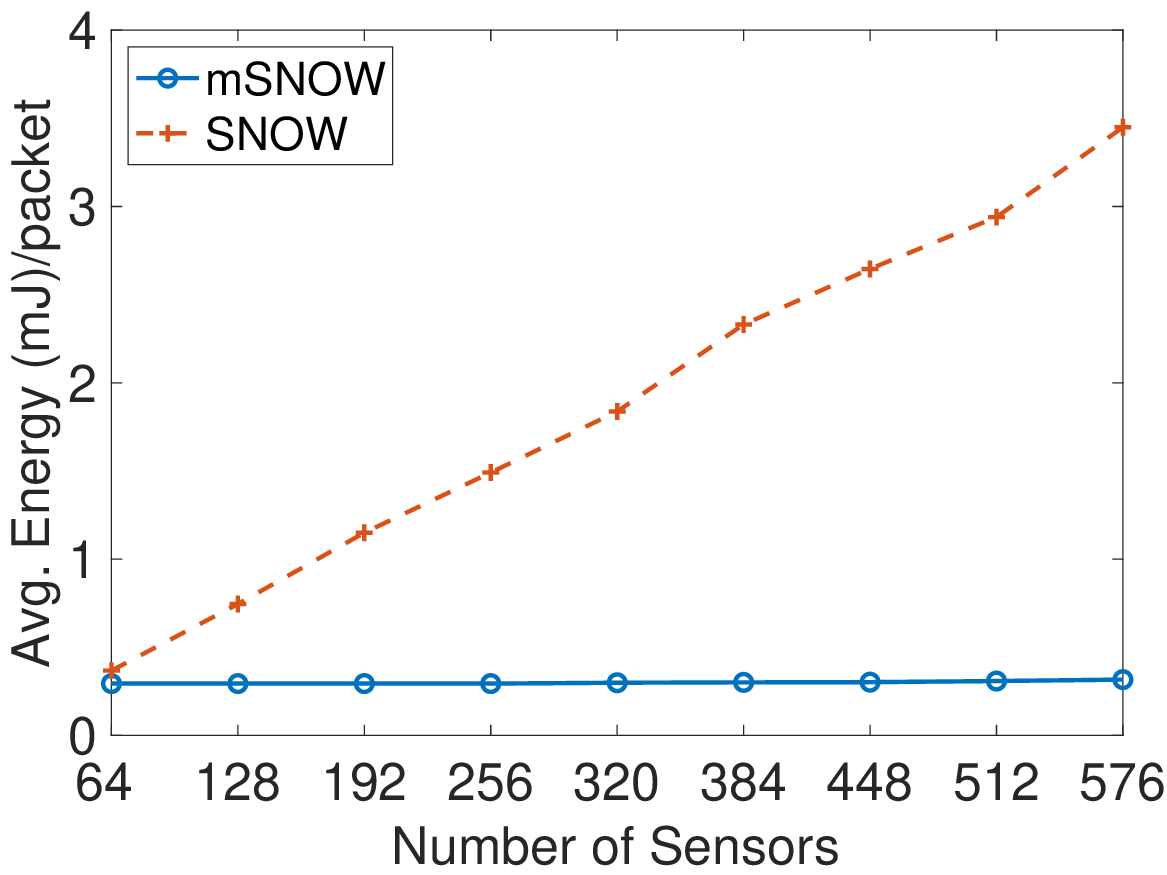}
      }
    \caption{Network performance evaluation in mSNOW uplink communications.} 
    \label{fig:networkperform}
\end{figure*}
\subsection{Evaluating Network Performance}
\subsubsection{During Uplink Communications in mSNOW}
In this section, we evaluate several network parameters in mSNOW uplink communications, including throughput (kbps), average latency per packet in data collection, and average energy consumption per packet at the sensors. Additionally, we compare our network performance with the existing SNOW design (MAC-enabled), as described in Section~\ref{sec:snowdesign}. In this setup, we use all 64 subcarriers in the 547 MHz -- 560 MHz band. As noted earlier in Section~\ref{sec:eval_setup}, all the sensors using odd- and even-numbered subcarriers get PN sequences from the sets $PNs_1$ and $PNs_2$, respectively. In this evaluation, we create a {\em convergecast} scenario and analyze the aforementioned network parameters, where each sensor transmits 100 40-byte packets (including 12-byte headers) with a random inter-packet interval between 0 to 3 ms.

\noindent{\bf Throughput.} In our evaluation, we consider the overall network throughput to be the overall effective bitrate (excluding the 12-byte packets' headers) at the BS in our convergecast scenario. In this setup, we consider various numbers of sensors up to $9 \times 64 = 576$. Figure~\ref{fig:nthroughput} shows the overall network throughput in mSNOW when numerous sensors between 64 and 576 transmit concurrently using 64 subcarriers, each subcarrier having a minimum and a maximum of 1 and 9 sensors, respectively. As shown in this figure, mSNOW achieves an overall bitrate of approximately 2.56 Mbps (compared to approximately 2.04 Mbps in SNOW) and 5.03 Mbps (compared to approximately 2.01 Mbps in SNOW) when 64 and 128 sensors transmit concurrently. As we increase the number of sensors, our throughput increases almost linearly, unlike the fixed or slightly decreasing throughput in the existing SNOW as it can decode concurrently from 64 sensors at any given instance. For example, our overall bitrate is approximately 20.31 Mbps (vs. approximately 1.94 Mbps in existing SNOW) when 576 sensors transmit concurrently. mSNOW thus has an approximately 9x throughput compared to SNOW.

\noindent{\bf Latency.}
Figure~\ref{fig:nlatency} shows the average latency per packet in convergecast while taking into account the lost packets (without using ACK but by a curve fitting approach so that we can emulate a 100\% reliability) as we increase from 64 to 576 sensors. Latency refers to the time it takes for a packet to be correctly delivered at the BS. As shown in this figure, the average (across the sensors) per-packet latency is approximately 5.6 ms when 64 sensors transmit concurrently in mSNOW. As we increase the number of sensors up to 576, the average per-packet latency stays in the range 5.6 -- 6.03 ms due to the massive concurrency in mSNOW. In contrary, the average per-packet latency increases linearly or at a higher rate as we increase the number of sensors from 64 to 576 in the existing SNOW, which is due to its CSMA-based MAC protocol. This simulation thus confirms the timeliness in our design, which may help many time-critical or real-time IoT or CPS applications.

\noindent{\bf Energy Consumption.}
Figure~\ref{fig:nenergy} depicts the average (across the sensors) per-packet energy consumption (for 100\% reliability) in our convergecast scenario. We calculate the energy based on the energy model of CC1310 transmitter (Tx current: 17.5 mA, idle current: 0.5 mA, and sleep current: 0.2$\mu$A at 0 dBm transmission power) that can operate in TV white spaces~\cite{saifullah2016snow}. As shown in Figure~\ref{fig:nenergy}, our average per-packet energy consumption stays almost the same (in the approximate range 0.2940 -- 0.3166 mJ) when we transmit concurrently and increase the number of sensors from 64 to 576. The average per-packet energy consumption in the existing SNOW increases linearly or at a higher rate (in the approximate range 0.3675 -- 3.4493 mJ) as the number of sensors is increased from 64 to 576. This is due to limited concurrency in SNOW (compared to mSNOW) and its CSMA-based MAC protocol. Our design thus shows better energy efficiency at the sensors, which may improve the lifetime of remote IoT or CPS applications.
\revise{Overall, our evaluations of various network parameters indicate that mSNOW allows for much more scalability in uplink communications than SNOW, which may enable massive scalability, timeliness in data collection, and greater sensor lifetime in IoT and CPS applications.}

\subsubsection{Network Performance in Downlink Communications}
\revise{For evaluating the performance in downlink communications, we consider a scenario of parallel peer-to-peer (P2P) communications in mSNOW. P2P communications are usually very common in multi-hop wireless senor-actuator networks (WSANs); however, are typically not supported in the LPWANs, which may be limiting for the emerging IoT and CPS applications~\cite{survey_lpwan, survey_tvws}. Note that the sensors in mSNOW (as well as in SNOW) are asynchronous and cannot communicate directly with each other. The P2P communications in mSNOW is thus enabled by the BS (e.g., similar to the controller/gateway WSANs), where a sender (e.g., a sensor node) sends the data to the BS and the BS forwards that data to the intended receiver (e.g., an actuator node). Such P2P communications thus invoke the downlink communication capability of the mSNOW BS and are our focus in this evaluation.}

\begin{figure}[!htbp]
    \centering 
    \includegraphics[width=0.3\textwidth]{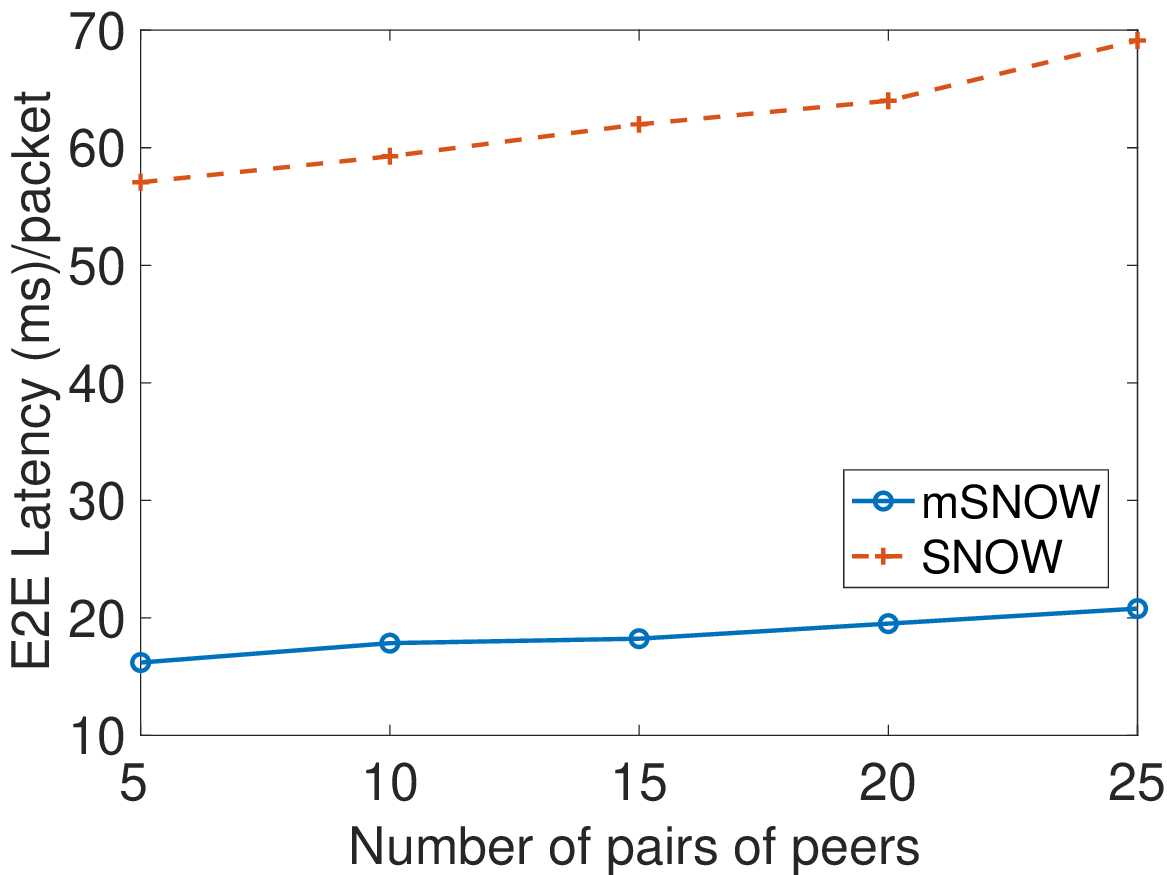}
    \caption{End-to-End latency in P2P communications.}
    \label{fig:dlink_latency}
\end{figure}
\revise{In this setup, we generate different numbers of pairs of peers, up to 25 (since P2P communications are less common scenarios) in mSNOW with the following properties to show a performance improvement over SNOW. (i) Senders and receivers in the pairs transmit on and receive from two different sets of D-OFDM subcarriers, (ii) each subcarrier used by a sender has no less than 5 senders in total concurrently transmitting to their peers, (iii) a subcarrier used by a receiver has no less than 5 receivers concurrently receiving from their peers, and (iv) the BS Tx-Radio may transmit pending packets to the receivers while the Rx-Radio may receive new packets from the senders.
With the above requirements, we allow each sender in each pair to transmit consecutive 1000 40-byte packets to their peers. We also repeat this setup 100 times with randomly generated sets of subcarriers (both for sending and receiving) from the available 64 subcarriers in our evaluation.}

\revise{To compare the performance in downlink communications between mSNOW and SNOW, we choose the {\em end-to-end} (E2E) network latency as a metric, which is defined as the total delay for a sender to successfully deliver (via the BS) a packet to a receiver in a pair of peers. Figure~\ref{fig:dlink_latency} depicts the average E2E latency per packet in mSNOW and compares it with SNOW as we vary the number of pairs of sensor peers between 5 and 25. As shown in this figure, when 5 pairs are active in the network, the average E2E latency per packet is approximately 16.21 ms in mSNOW (compared to 57.06 ms in SNOW). In a nutshell, mSNOW observes less E2E latency due to its ability to concurrently receive from or concurrently transmit on a single subcarrier, while SNOW depends on its CSMA-based MAC protocol or round-robin scheduling, respectively. Figure~\ref{fig:dlink_latency} also shows that the E2E latencies per packet in SNOW increase at a slightly higher rate than mSNOW as we enable more parallel pairs of peers in the network. For example, in the case of 25 parallel pairs of peers in the network, mSNOW and SNOW observe average E2E latencies of 20.79 ms and 69.13 ms per packet, respectively, demonstrating the better suitability of mSNOW.}

\revise{Overall, our evaluations on both uplink and downlink communications demonstrate that the design improvements in mSNOW over SNOW are significant, which may enable the emerging IoT and CPS applications that require tens of thousands of sensors with longer battery life while also making data-driven, time-sensitive decisions. In the following, we now conclude our paper.}


\section{Conclusions}\label{sec:conclusion}
\revise{In this paper, we have proposed mSNOW by significantly advancing the PHY layer of an LPWAN technology called SNOW, which has enabled unprecedented concurrency in the LPWAN's design. In the process of enabling massive concurrency, and hence scalability, in mSNOW, we have developed a set of PN sequences based on Gold code, which causes minimal interference within and across the mSNOW D-OFDM subcarriers when used by numerous asynchronous and concurrently transmitting sensors in both uplink and downlink communications. Our evaluation results have shown that we have achieved approximately 9x more scalability in mSNOW compared to SNOW. Our evaluation results have also suggested that mSNOW significantly improved the per-packet latency and energy consumption in the sensors at the network level. Overall, our design may motivate massive concurrency in communications in general LPWANs and WSNs through its innovations and open-source implementation, thereby encouraging the next generation of IoT and CPS applications requiring tens of thousands of sensors.}

\section*{Acknowledgements}
This work was supported by CUNY Queens College through RFCUNY grant \#90922-02 09 and Google Cyber NYC Institutional Research through RFCUNY grant \#7E263-07 02.

\bibliographystyle{IEEEtran}
\balance
\bibliography{IEEEabrv,snow}
\end{document}